\documentclass[referee]{raa}            

\usepackage{graphicx,times}             
\input{epsf.sty}                        
\input{psfig.sty}                       

\begin{document}

   \title{Astrophysical parameters of ten poorly studied open star clusters.
}

   \volnopage{Vol.0 (200x) No.0, 000--000}      
   \setcounter{page}{1}          

   \author{Tadross, A. L.
      \inst{1}
  \and R. El-Bendary
    \inst{1}
     \and A. Osman
      \inst{1}
       \and N. Ismail
      \inst{2}
       \and A. Bakry
      \inst{2}
   }

   \institute{National Research Institute of Astronomy \& Geophysics, Cairo, Egypt.; {\it altadross@yahoo.com}\\
        \and
             Faculty of Science, Al-Azhar University, Cairo, Egypt.\\
   }

   \date{Received~~2009 month day; accepted~~2009~~month day}

\abstract{We present here the fundamental parameters of ten open star clusters, nominated from Kronberger et al. (2006) who presented some new discovered stellar groups on the basis of 2MASS photometry and DSS visual images. Star counts and photometric parameters (radius, membership, distances, color excess, age, luminosity function, mass function, total mass, and the dynamical relaxation time) have been determined for these ten clusters for the first time. In order to calibrate our procedures, the main parameters (distance, age, and color excesses) have been re-estimated for another five clusters, which are studied by Kronberger et al. (2006) as well.
}

   \authorrunning{Tadross, A. L. et al.}            
   \titlerunning{Astrophysical parameters of ten poorly studied open star clusters. }  

   \maketitle

%
%
\section{Introduction}
Open star clusters (OCs) are very important tools in studying the formation and evolution of the Galactic disk and the stellar evolution. The fundamental physical parameters of OCs, e.g. distance, reddening, age, and metallicity are necessary for studying the Galactic disk. The Galactic, radial and vertical, abundance gradient also can be studied by OCs (Hou et al. 2000, Chen et al. 2003, Kim \& Sung 2003, Tadross 2003, Kim et al. 2005). They are excellent probes of the Galactic disc structure (Janes \& Phelps 1994, Bica et al. 2006). The strong interest of OCs results come from their fundamental properties. Among the 1784 currently known OCs, more than half of them have been poorly studied or even unstudied up to now, Piatti et al. (2011). The current paper is thus part of our continuation series whose goal is to obtain the main astrophysical properties of previously unstudied OCs using modern databases. 73\% of the clusters under investigation (11 of 15) are lying very near to the Galactic plane (-2.5$^{o}$ $\leq$ b $\leq$ +4.2$^{o}$). The most important thing for using NIR database (2MASS) is the powerful detection of the star clusters behind the hydrogen thick clouds, which are concentrated on the Galactic plane. Kronberger et al. (2006); hereafter K06; presented a list of stellar groups, which are not confirmed if they are open clusters or not. Table 1 contains the equatorial and Galactic coordinates of all the fifteen clusters which are nominated for the present work. Five of them have their basic parameters, which were estimated by K06. These ones have been used as calibrated clusters for our procedures; they signed by small letter (c) in Table 1. The photometric and structure properties for the remaining ten clusters have been studied here for the first time.

This paper is organized as follows. In sect. 2, the data reductions have been presented; Star counts and Field-star decontamination are presented in sect. 3 and sect. 4 respectively. In sect. 5, the photometric analysis CMDs is illustrated. The calibration of this work is obtained in sect. 6. The luminosity and mass functions are presented in sect. 7. Dynamical state is introduced in sect. 8.  Finally, the conclusion is devoted to sect. 9.

\begin{table*}
\centering
\caption{Equatorial \& Galactic positions of the clusters under investigation; sorted by right ascensions.}
\centerline{
\begin{tabular}{ccrrrr}
\hline
            Cluster
            & $\alpha~^{h}~^{m}~^{s}$
            & $\delta~^{\circ}~{'}~{''}$
            & G. Long.$^{\circ}$
            & G. Lat.$^{\circ}$\\
\hline
 Riddle 4~~~~~~~~        &  02:07:22.7  &  +60:15:25 & 132.222 & --1.238 \\
 Juchert 9 c~~~~~     &  03:55:21.0  &  +58:23:30 & 58.838  & 58.392 \\
 Kronberger 1 c  &  05:28:21.0  &  +34:46:30 & 82.088  & 34.775 \\
 Teutsch 51 c~~~~    &  05:53:51.9  &  +26:49:47 & 88.466  & 26.829 \\
 Teutsch 11~~~~~~      &  06:25:24.4  &  +13:51:59 & 197.654 & 0.650 \\
 Alessi 53~~~~~~~~~       &  06:29:24.5  &  +09:10:39 & 202.261 & --0.664 \\
 Alessi 15~~~~~~~~~       &  06:43:04.0  &  +01:40:19 & 210.489 & --1.099 \\
 Juchert 12~~~~~~~      &  07:20:56.7  &  --22:52:00 & 236.561 & --4.123 \\
 Riddle 15~~~~~~~~       &  19:11:33.0  &  +14:50:04 & 48.357  & 2.455 \\
 Juchert 1~~~~~~~~~       &  19:22:32.0  &  +12:40:00 & 47.728  & --1.00 \\
 Patchick 89~~~~~~     &  19:59:33.0  &  +49:18:45 & 83.644  & 10.129 \\
 Toepler 1~~~~~~~~~       &  20:01:17.6  &  +33:36:54 & 70.30   & 1.719 \\
 ADS 13292 c~~~~~     &  20:02:23.3  &  +35:18:41 & 71.861  & 2.425 \\
 Teutsch 30 c~~~~~    &  20:27:43.0  &  +36:04:32 & 75.354  & --1.425 \\
 Teutsch 144~~~~~~     &  21:21:43.9  &  +50:36:36 & 92.735  & 0.459 \\
\hline\\
\end{tabular}}
$c$) Calibrated clusters; those have the main parameters (ages, distances, and color excesses) estimated by K06.
\end{table*}

\section{Data Reductions}

The astrophysical parameters of the investigated clusters are used J, H, and Ks photometry obtained from Two Micron All Sky Survey (2MASS) of Skrutskie et al. (2006), available at (www.ipac.caltech.edu/2mass/releases/allsky).

2MASS is designed to close the gap between our current technical capability and our knowledge of the NIR Sky. It provides J, H, and Ks band photometry for millions of galaxies and nearly a half-billion stars (Carpenter 2001). While the 2MASS data alone can give important contributions to many fields of study, the scientific impact of many programs can be further enhanced by comparing 2MASS photometry with existing photometric measurements or by conducting follow-up observations.
Two 1.3-m telescopes are used, one at Mount Hopkins in Arizona (Northern Survey) and the other at Cerro Tololo in Chile (Southern Survey). Each telescope was equipped with three-channel camera, each channel consisting of 256 x 256 array HgCdTe detectors. It is uniformly scanning the entire sky in the three NIR bands J (1.25 $\mu$m), H (1.65 $\mu$m) and Ks (2.17 $\mu$m). This survey has proven to be a powerful tool in the analysis of the structure and stellar content of open clusters (Bica et al. 2003, Bonatto \& Bica 2003). From Soares \& Bica (2002), we can see that the errors are more affected for Ks band at a given magnitude. So, J \& H data have been used here to probe the faint stars of these clusters with more accuracy.\\

Data extraction has been performed using the known tool of VizieR for 2MASS database. The investigated clusters' data have been downloaded under the following conditions:\\
\\
$\circ$	The clusters' data extracted at a preliminary radius of about 20 arcmin from their obtained centers, and a nearby control field of the same area should be downloaded as well;\\
$\circ$	The clusters should have good images on the Digitized Sky Survey (DSS);\\
$\circ$	They should be apparently rich clusters with prominent sequences in their CMDs;\\
$\circ$	The photometric completeness limit at J $<$ 16.5 mag is applied on the 2MASS data to avoid the over-sampling at the lower parts of their CMDs (cf. Bonatto et al. 2004);\\
$\circ$	Stars with observational uncertainties in {J, H, Ks} $>$ 0.20 mag have been eliminated;\\
$\circ$	Membership criteria are adopted for the location of the stars in the CMD curves within 0.10 - 0.15 mag around the ZAMS, Claria \& Lapasset (1986).\\

\section{Star Counts}
\subsection{Cluster center's determination}
A cluster's center is defined as the center of the cluster's mass or the location of the maximum stellar density (the number of stars per unit area in the direction of the cluster). The cluster center ($\alpha$, $\delta$) has been estimated as explained in details in Tadross (2005). All the clusters' centers in the present work are estimated and found in a good agreement with K06 within errors of only few arc seconds.

\subsection{Clusters' diameters determination}
A cluster real diameter can be determined using the radial surface density of the stars of the cluster. The cluster border is defined as the surface which covers the entire cluster area and reaches enough stability in the background density, i.e. the difference between the observed density profile and the background one is almost equal zero; for more details see Tadross (2005). It is noted that the determination of a cluster radius is made by the spatial coverage and uniformity of 2MASS photometry which allows one to obtain reliable data on the projected distribution of stars for large extensions around the clusters' center, Bonatto et al. (2005). Although the spatial shape of the cluster may not be perfectly spherical, the fitting of King (1962) model has been applied to derive the cluster limited radius and the core radius as well. Fig. 1 represents an example to derive the cluster's radius of Patchick 89. On the other hand, knowing the cluster's total mass (Sec. 7), the tidal radius can be given by applying the equation of Jeffries et al. (2001):
\begin{equation}
R_{t} = 1.46 ~~ (M_{c})^{1/3}
\end{equation}
where $R_{t}$ and $M_{c}$ are the tidal radius and total mass of the cluster respectively.

\section{Field-star decontamination}
Usually, field stars contaminate the CMDs of a cluster, particularly at faint magnitudes and red colors. Most clusters are located near the disk or/and the bulge of the Galaxy and shows crowded contaminated main sequences. These contaminated stars are always seen as a vertical redder sequence parallel to the cluster's main sequence. CMDs of such clusters surely contain field stars that might lead to artificial isochrone solutions, i.e. one can always "fit" isochrones to such CMDs. Therefore, field-star decontamination must be used to define the clean CMDs and get better isochrone fitting. To achieve this, where at low latitudes the field stellar population is not homogeneous, a ring around the cluster with a radius four times the cluster limited radius is used as a nearby control field. From the comparison of the CMDs of such a cluster and its control field for a given magnitude and color range, we have counted the number of stars in a control field and subtracted this number from the cluster's CMDs. It can be noted that, for a good separated cluster, the mean density of the control field is always less than the central region of the cluster.

\begin{figure}
\begin{center}
{\includegraphics[width=7cm]{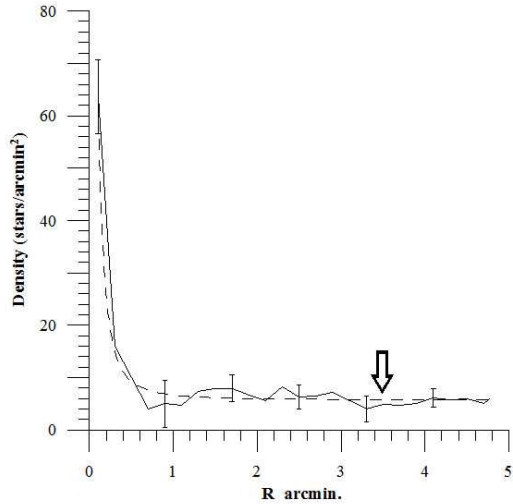}}
\end{center}
\caption{An example for the radius determination of "Patchick 89", the curved dashed line represents the fitting of King (1962) model. The arrow refers to the point at which the radius is taken (at 3.5 arc min). While the error bars represent 1 $\sigma$ Poisson scatter around the King fit. The background field density = 5.7 stars per arcmin$^{2}$, and the core radius = 0.07 arc min.}
\end{figure}

\section{Photometric analysis (CMDs)}
CMDs are established for the stars inside radii of 1$^{'}$, 2$^{'}$, 3$^{'}$ etc. from the optimized coordinates of the centers of the clusters under investigation. We have fitted the new theoretical isochrones computed with the 2MASS $J, H$ and $K_S$ filters (Bonatto et al. 2004, Bica et al. 2006) to derive the cluster parameters. The simultaneous fittings were attempted on the $J\sim(J-H)$ and $K_S\sim(J-K_S)$ diagrams for the inner stars, which should be less contaminated by the background field.
If the number of stars are not enough for an accepted fitting, the next larger area is included, and so on. In this way, different isochrones of solar metallicity (Z = 0.019) of different ages have been applied on the CMDs of each cluster, fitting the lower envelope of the points matching the main sequence stars, turn-off point and red giant positions. Solar metallicity isochrones have been used with all our investigated clusters to unify the estimation of the main parameters, especially ages, distances and reddening. JHK color-color diagrams (CCD) do not strongly depend on metallicity whereas reddening is represented as a straight vector on CCD. On the other hand, for relatively young clusters, the isochrone fitting could be somewhat tricky, since different pairs of distance modulus and color excess could satisfactorily fit the cluster ZAMS. In order to avoid such degeneracy, CCD has been built for each cluster and thus,  guiding by the Galactic reddening values of Schlegel et al. (1998), the realistic values of $E_{J-H}$ and $E_{J-K_S}$ can be estimated as shown in Fig. 2. Although Schlegel's reddening values are often overestimated at low Galactic latitudes, it is still a useful source of data. Comparing our estimated reddening values with the reliable ones of Schlegel's, we found that our sample are in agreement with Schlegel's values within ranging errors of 0.08 mag. The distance modulus is taken at the proper values within a ranging fitting error of about $\pm 0.10$ mag.

The observed data has been corrected for interstellar reddening using the coefficients ratios $\frac {A_{J}}{A_{V}}= 0.276$ and $\frac {A_{H}}{A_{V}}= 0.176$, which were derived from absorption rations in Schlegel et al. (1998), while the ratio $\frac {A_{K_S}}{A_{V}}= 0.118$ was derived from Dutra et al. (2002).
Therefore $\frac {E_{J-H}}{E_{B-V}}= 0.309$, $\frac {E_{J-K_S}}{E_{B-V}}= 0.488$, and then $\frac {E_{J-K_S}}{E_{J-H}}\approx$ 1.6 can be derived easily from the above ratios, where R$_{V}=\frac {A_{V}}{E_{B-V}}= 3.1$.

Once the distance from the Sun ($R_{\odot}$) is estimated, the distance from the galactic center ($R_{G}$) and the projected distances on the galactic plane from the Sun ($X_{\odot}~\&~Y_{\odot}$) and the distance from galactic plane ($Z_{\odot}$) can be determined. For more details about the estimation of the Galactic geometric distances see Tadross (2011).

\begin{figure*}
\begin{center}
{\includegraphics[width=11cm]{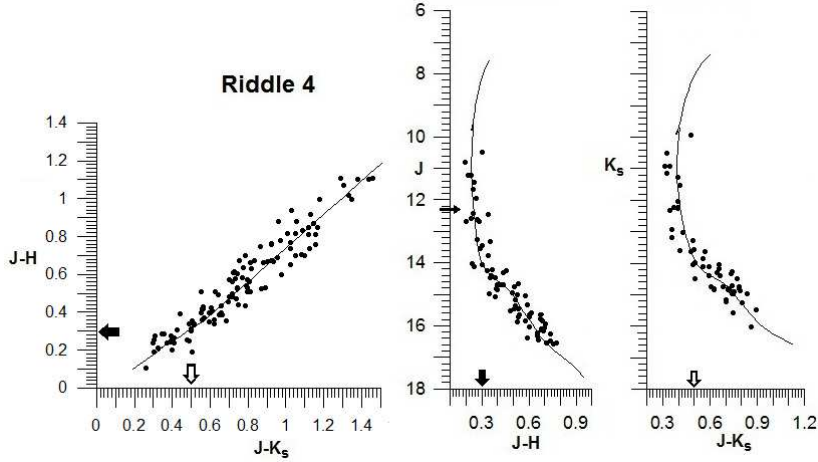}}
\end{center}
\caption{JHK color-color and color-magnitude diagrams for the members of Riddle 4, as an example for the clusters under investigation.}
\end{figure*}

\section{Calibration of the present work}

The main parameters of known previously studied the five clusters (Juchert 9, Kronberger 1, ADS 13292, Teutsch 30 and Teutsch 51) have been re-estimated in the present work and compared with those taken from K06. Noted that Kronberger 1 and Teutsch 51 are taken from Koposov et al. (2008).

In this context, the main parameters of Kronberger 1 have been compared with Dias \& Webda catalogs, K06 and Koposov et al. (2008). We found that our results agree with those of Dias \& Webda and K06 more than those of Koposov et al. (2008). On the other hand, Teutsch 51, was published earlier as Koposov 52 and KSE18 (Koposov et al. 2005, Zolotukhin et al. 2006), and then by K06. The main results of Teutsch 51 are found agree somewhat with those of Koposov et al. (2008) more than those of K06. However, Fig. 3 represents comparisons of the present main results (distance moduli, ages and color excesses) with the previous ones of K06.
It is clearing that the derived parameters are reasonable and very close to the published ones, which indicates that our reduction procedure in using 2MASS database is very reliable.

\begin{figure*}
\begin{center}
{\includegraphics[width=14cm]{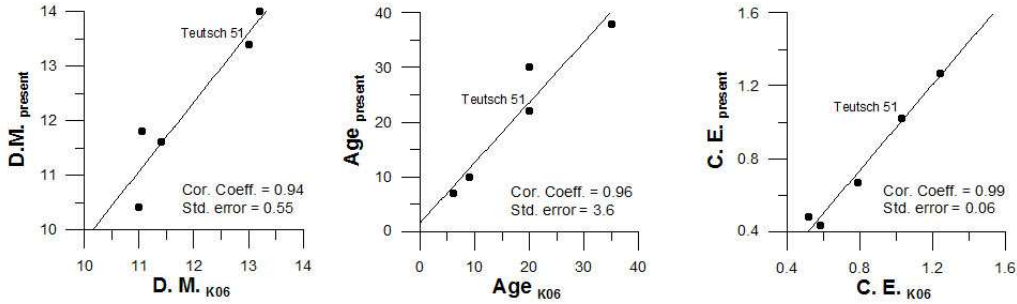}}
\end{center}
\caption{Comparing our main results (distance moduli, ages, and color excesses) of the five clusters with those of K06 (the parameters of Teutsch 51 are taken from Koposov et al. (2008)). The correlation coefficients and standard errors are shown for each relation respectively.}
\end{figure*}

\section{Luminosity and Mass Functions}

The luminosity function (LF) gives the number of stars per luminosity interval, or in other words, the number of stars in each magnitude bin of the cluster. It is used to study the properties of large groups or classes of objects, such as the stars in clusters or the galaxies in the Local Group.

In order to estimate the luminosity function, firstly we should count the observed stars in terms of absolute magnitude after applying the distance modulus. The magnitude bin intervals are selected to include a reasonable number of stars in each bin and for the best possible statistics of the luminosity and mass functions. From LF, we can infer that massive bright stars seem to be centrally concentrated more than low masses and fainter ones (Montgomery et al. 1993).

As known the LF and mass function (MF) are correlated to each other according to the Mass-luminosity relation. The accurate determination of both of them (LF \& MF) suffers from some problems e.g. the field contamination of cluster members and the observed incompleteness at low-luminosity (or low-mass) stars, which may affect even poorly populated, relatively young clusters (Scalo 1998).
On the other hand, the properties and evolution of a star are closely related to its mass, so the determination of the initial mass function (IMF) is needed, that is an important diagnostic tool for studying large quantities of star clusters. IMF is an empirical relation that describes the mass distribution (a histogram of stellar masses) of a population of stars in terms of their theoretical initial mass (the mass they were formed with). Scalo (1998) defined the IMF in terms of a power law as following:
\begin{equation}
\large \frac{dN}{dM} \propto M^{-\alpha}
\end{equation}

where $\frac{dN}{dM}$ is the number of stars of mass interval ($M:M+dM$) within a specified volume of space, and $\alpha$ is a dimensionless exponent. The IMF for massive stars ($>$ 1 $M_{\odot}$) has been studied and well established by Salpeter (1955), where $\alpha$ = 2.35. This form of Salpeter shows that the number of stars in each mass range decreases rapidly with increasing mass, see Fig. 4.

The mass of each star in the investigated clusters has been estimated from a polynomial equation developed from the data of the solar metal abundance isochrones (absolute magnitudes vs. actual masses) at a specific age of each cluster individually. The summation of multiplying the number of stars in each bin by the mean mass of that bin yields the total mass of each cluster.

\begin{figure}
\begin{center}
{\includegraphics[width=8cm]{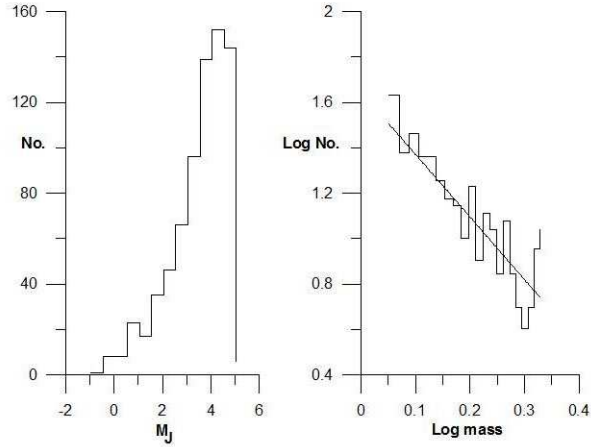}}
\end{center}
\caption{An example for luminosity and mass functions of "Teutsch 144".}
\end{figure}

\section{Dynamical state}

The relaxation time (T$_{R}$) of a cluster is defined as the time in which the cluster needs from the very beginning to build itself and reach the stability state against the contraction and destruction forces, e.g. gas pressure, turbulence, rotation, and the magnetic field (cf. Tadross 2005).  T$_{R}$ is depending mainly on the number of members and the cluster radius and the mean total mass. It can be given in the next equation:
\begin{equation}
\large T_{R} = \frac{8.9 \times 10^{5} \sqrt{N} \times R_{h}^{1.5}}{\sqrt{m} \times log (0.4 N)}
\end{equation}
where $N$ is the number of the cluster members, $R_{h}$ is the radius containing half of the cluster mass in parsecs, and m is the average mass of the cluster in solar unit (Spitzer \& Hart 1971). Using the above equation we can estimate the dynamical relaxation time of each cluster and compare it with its age. If the cluster's age is found greater than its relaxation time, then the cluster was dynamically relaxed, and vice versa.

\section{Conclusions}
Our analysis has been applied for estimating the astrophysical parameters of some new discovered stellar groups according to K06. Hence, ten really open star clusters have been detected, which have real stellar density profiles with IMF slopes around the Salpeter's (1955) value. Also, the ages of these clusters are found to be greater than their relaxation times that infer that these clusters are dynamically relaxed indeed.

All parameters of the investigated clusters are listed in Table 2, and the CMDs of each cluster are shown in Fig. 5.

\begin{acknowledgements}
It is worthy to mention that, this publication made use of WEBDA and the data products from the Two Micron All Sky Survey (2MASS), which is a joint project of the University of Massachusetts and the Infrared Processing and Analysis Center/California Institute of Technology, funded by the National Aeronautics and Space Administration and the National Science Foundation (NASA). Catalogues from CDS/SIMBAD (Strasbourg) and Digitized Sky Survey DSS images from the Space Telescope Science Institute have employed.
\end{acknowledgements}

\begin{figure*}
\begin{center}
{\includegraphics[width=14cm]{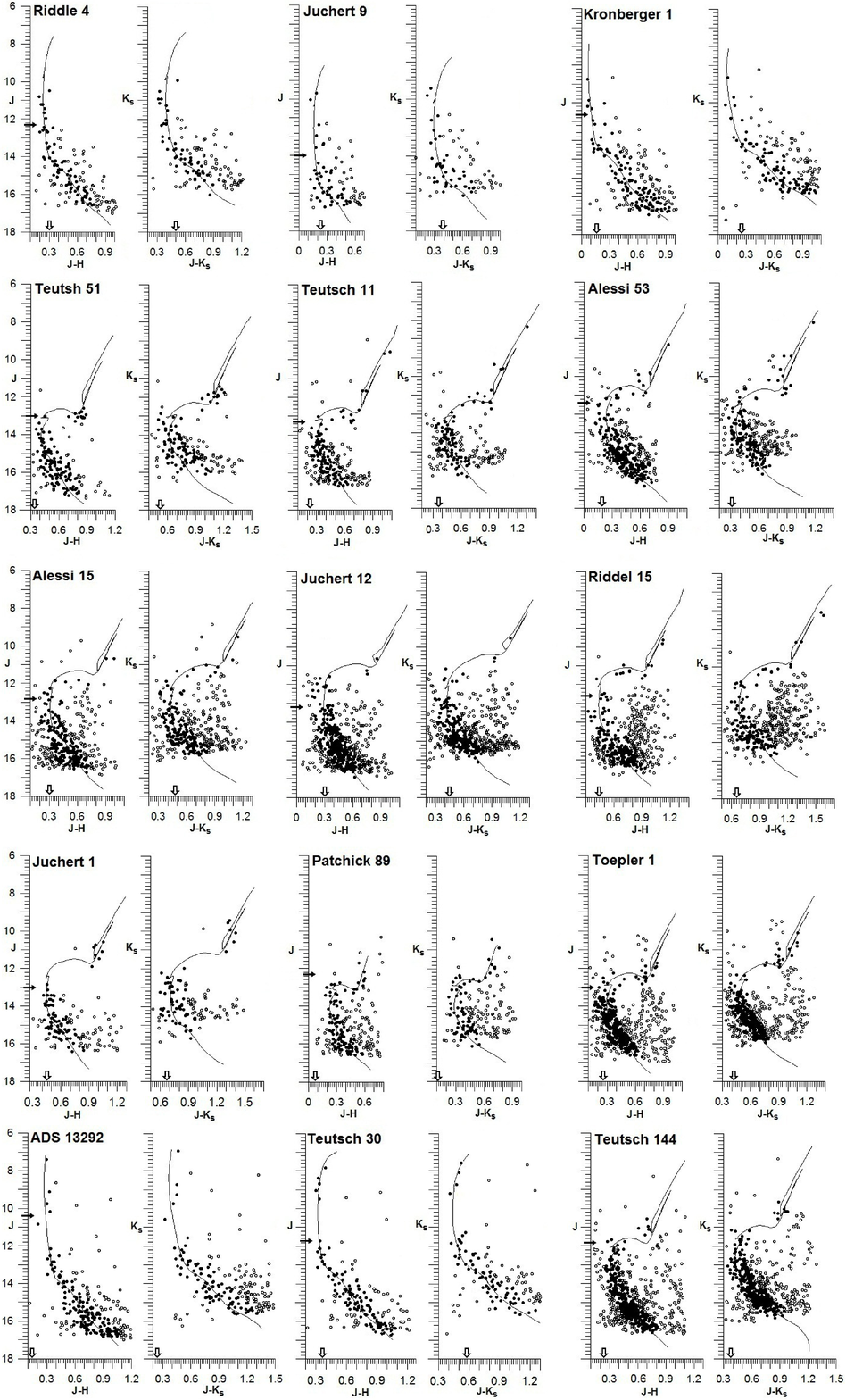}}
\end{center}
\caption{Color-magnitude diagrams of the clusters under investigation. Filled and empty circles denote the stars lying closely and far from the fitted isochrones (photometric probable members and field stars) respectively. Horizontal and vertical arrows refer to the estimated distance modulus and the values of the color excesses in both diagrams for each cluster respectively.}
\end{figure*}

\begin{table*}
\caption{The derived astrophysical parameters for the investigated clusters. Columns display, respectively, cluster name, age, distance modulus, distance from the sun, reddening values, angular radius, core radius, linear radius, tidal radius, distance from the Galactic center, the projected distances on the Galactic plane from the sun, X \& Y, the distance from Galactic plane, total mass, membership number, and time of relaxation.}
\centerline{
\begin{tabular}{ccrrrrrrrrrrrrrr}\\
\hline
            Cluster~~~~~ & Age  & m-M  & $R_{\odot}$~~~~~~ & E$_{B-V}$ & R  & R$_{c}$ &  R$_{L}$ & R$_{t}$ & $R_{G}$
            & $X_{\odot}$  & $Y_{\odot}$  & $Z_{\odot}$ & Mass & Mem. & T$_{R}$ \\
                        & \it Myr & \it mag & \it pc~~~~~~  & \it mag  & \it $^{\prime}$  & \it $^{\prime}$  & \it pc   & \it pc
            & \it kpc  & \it pc & \it pc & \it pc & \it M$_{\odot}$ &  & \it Myr \\
\hline\\
 Riddle 4~~~~~~~~         & 50    & 12.3 & 1993 $^{\pm 92}$   & 0.91  & 2.2 & 0.13 & 1.3 & 9  & 9.95  & 1339   & 1475 & --43 & 230 & 120 & 2.1 \\
 Juchert 9 c~~~~~      & 35    & 14.0 & 4694 $^{\pm 216}$  & 0.73  & 1.8 & 0.16 & 2.5 & 8  & 7.28  & --1273 & 2105 & 3998 & 175 & 65 & 4.2 \\
 Kronberger 1 c   & 20    & 11.6 & 1715 $^{\pm 79}$   & 0.48  & 2.2 & 0.30 & 1.1 & 8  & 8.44  & --194  & 1396 & 978 & 195 & 140 & 2.1 \\
 Teutsch 51 c~~~~     & 800   & 13.0 & 2618 $^{\pm 121}$  & 1.03  & 2.1 & 0.17 & 1.6 & 9  & 8.83  & --63   & 2336 & 1182 & 210 & 136 & 3.4 \\
 Teutsch 11~~~~~~   & 500   & 13.3 & 3443 $^{\pm 159}$  & 0.70  & 3.0 & 0.26 & 3.0 & 9  & 11.83 & 3280   & --1044 & 39 & 220 & 155 & 9.6 \\
 Alessi 53~~~~~~~~~        & 500   & 12.4 & 2360 $^{\pm 109}$  & 0.61  & 6.4 & 0.29 & 4.4 & 10 & 10.72 & 2184   & --894 & --27 & 365 & 255 & 19.3 \\
 Alessi 15~~~~~~~~~        & 450   & 12.8 & 2509 $^{\pm 116}$  & 0.91  & 6.0 & 0.12 & 4.4 & 13 & 10.74 & 2161   & --1273 & --48 & 735 & 495 & 22.9 \\
 Juchert 12~~~~~~~       & 300   & 13.2 & 3016 $^{\pm 139}$  & 0.91  & 5.0 & 0.12 & 4.4 & 9  & 10.47 & 1658   & --2510 & --217 & 250 & 125 & 13.5 \\
 Riddle 15~~~~~~~~        & 500   & 12.6 & 1925 $^{\pm 89}$   & 1.33  & 2.5 & 0.08 & 1.4 & 11 & 7.36  & --1278 & 1437 & 82 & 385 & 275 & 3.6 \\
 Juchert 1~~~~~~~~~        & 400   & 13.0 & 2286 $^{\pm 105}$  & 1.36  & 1.6 & 0.05 & 1.1 & 9  & 7.16  & --1538 & 1691 & --40 & 200 & 157 & 2.1 \\
 Patchick 89~~~~~~      & 1600  & 12.3 & 2646 $^{\pm 122}$  & 0.21  & 3.5 & 0.07 & 2.7 & 9  & 8.62  & --288  & 2588 & 465 & 220 & 155 & 8.1 \\
 Toepler 1~~~~~~~~~        & 400   & 13.0 & 2890 $^{\pm 133}$  & 0.79  & 4.0 & 0.10 & 3.4 & 15 & 8.00  & --974  & 2719 & 87 & 1120 & 668 & 16 \\
 ADS 13292 c~~~~~      & 10    & 10.4 & 1021 $^{\pm 47}$   & 0.42  & 1.5 & 0.16 & 0.4 & 15 & 8.24  & --318  & 969 & 43 & 995 & 78 & 0.2 \\
 Teutsch 30 c~~~~~     & 35    & 11.7 & 1387 $^{\pm 64}$   & 1.12  & 1.8 & 0.19 & 0.7 & 11 & 8.26  & --350  & 1341 & --34 & 455 & 111 & 0.6 \\
 Teutsch 144~~~~~~      & 800   & 11.8 & 1704 $^{\pm 79}$   & 0.73  & 5.0 & 0.30 & 2.5 & 14 & 8.75  & 81     & 1702 & 14 & 890 & 745 & 12.4 \\
\hline\\
\end{tabular}}
$c$) Calibrated clusters; see Table 1.
\end{table*}

\end{document}